\documentclass[a4paper, 12pt]{article}

\usepackage{arydshln}
\usepackage{graphics}
\usepackage{amsmath}
\usepackage[pdftex]{graphicx,color}
\usepackage{tocloft}
\usepackage{amsmath, amssymb}
\usepackage{color}
\usepackage[bookmarks=false,pdfstartview=FitH]{hyperref}
\usepackage[usenames,dvipsnames]{xcolor}
\usepackage[margin=1in]{geometry}

\numberwithin{equation}{section}

\newcommand{\superN}{\mathcal{N}}

\newcommand{\beq}{\begin{equation}}
\newcommand{\eeq}{\end{equation}}
\newcommand{\beqq}{\begin{equation*}}
\newcommand{\eeqq}{\end{equation*}}
\newcommand\beqa{\begin{eqnarray}}
\newcommand\eeqa{\end{eqnarray}}
\newcommand\beqaa{\begin{eqnarray*}}
\newcommand\eeqaa{\end{eqnarray*}}
\newcommand\bea{\begin{array}}
\newcommand\eea{\end{array}}
\newcommand{\M}{{\cal M}}

\renewcommand{\[}{\left[}
\renewcommand{\]}{\right]}
\renewcommand{\(}{\left(}
\renewcommand{\)}{\right)}

\begin{document}

\title{Algebraic Curve for a Cusped Wilson Line}
\author{Grigory Sizov$^{1}$\footnote{\tt{grigory.sizov@kcl.ac.uk}}, \, Saulius Valatka$^1$\footnote{\tt{saulius.valatka@kcl.ac.uk}}
\vspace{20pt}\\
\small $^1$ King's College London, Department of Mathematics, \\
\small The Strand, London WC2R 2LS, UK\\}

\date{}

\maketitle

\begin{abstract}
We consider the classical limit of the recently obtained exact result for the anomalous dimension of a cusped Wilson line with the insertion of an operator with $L$ units of R-charge at the cusp in planar ${\cal N}=4$ SYM.
The classical limit requires taking both the 't Hooft coupling and $L$ to infinity. Since the formula for the cusp anomalous dimension involves determinants of size proportional to $L$, the classical limit requires a matrix model reformulation of the result. We construct such matrix model-like representation and find the corresponding classical algebraic curve. Using this we find the classical value of the cusp anomalous dimension and the 1-loop correction to it. We check our results against the energy of the classical solution and numerically by extrapolating from the quantum regime of finite L.
\end{abstract}

\clearpage


\section{Introduction}

The duality between $\superN=4$ super Yang-Mills theory in four dimensions and type IIB superstring theory on $AdS_5 \times S^5$ is undoubtedly the most studied and best understood example of the AdS/CFT correspondence \cite{AdSCFT}. The theories on both sides of the duality are also known to be integrable, meaning that one can hope to access non-perturbative regimes in both of them, which is highly non-trivial to do using conventional tools and methods (see the seminal paper \cite{Minahan:2002ve}, also see \cite{Beisert:2010jr} for a recent review of integrability in AdS/CFT).

Non-perturbative calculations in gauge theories are rare in general, yet they are very important for better understanding these theories. In particular, they allow us to better understand the AdS/CFT correspondence, which is a strong/weak duality. One such recent non-perturbative result in planar ${\cal N}=4$ SYM is the calculation of the anomalous dimension of a cusped Wilson line at any coupling which was done in \cite{Correa:2012at},\cite{Gromov:2012eu}, \cite{Gromov:2013qga}. The observable in question is a cusped Wilson line with the cusp angle $\phi$ and an angle $\theta$ regulating the coupling to the $\superN=4$ SYM scalars (see figure \ref{fig:WL}). In addition there is a scalar operator of R-charge $L$ inserted at the cusp. The calculation led to the formula for the cusp anomalous dimension $\Gamma_{L}$ which involved determinants of the size proportional to $L$. The result was subject to different kinds of tests, including weak and strong coupling expansions. While it is easy to perform a weak coupling expansion of the result, the strong coupling is not so straightforward. The problem is that the classical limit, in which we can compare the cusp anomalous dimension with the energy of the classical string, corresponds
to taking both the 't Hooft coupling $\lambda$ and the R-charge $L$ to infinity, while keeping $L/\sqrt{\lambda}$ finite. Since $L$ determines the size of the determinants in the formula for $\Gamma_{L}$, we can't directly take the large $L$ limit. The solution to this problem is to reformulate the formula as an expectation value in some matrix model. Then the classical value of $\Gamma_{L}$ will be given by the saddle-point approximation of a matrix integral. An elegant way to describe the solution in the classical limit is the algebraic curve method \cite{Beisert:2005bm},\cite{Gromov:2007aq},\cite{Kazakov:2004qf},\cite{Bargheer:2008kj} which we adopt in this paper. The algebraic curve in question was found in
\cite{Gromov:2012eu} for the limit $\theta=0,\;\phi\ll 1$ and here we generalize that construction to the case of arbitrary $\theta$ and $\phi\approx\theta$.

The paper is organized as follows. We start in section \ref{sec:Cusp} by reviewing the results of \cite{Gromov:2013qga}, which will be the starting point for this work. Then  in section \ref{sec:Matrix} we reformulate the problem in the language of matrix models, showing how the cusp anomalous dimension can be expressed as an expectation value in the aforementioned matrix model. In section \ref{sec:Classical} we review the corresponding classical string solution and in section \ref{sec:Theenergy} we find the algebraic curve and using it derive the classical energy and the 1-loop correction to it. We show that our results indeed agree with the known classical expansions for the cusp anomalous dimension. In section \ref{sec:Conclusions} we conclude by discussing our results and also the possible directions for continuing this work.


\section{Cusp anomalous dimension of a Wilson line}
\label{sec:Cusp}

\begin{figure}[t]
	\centering
		\includegraphics[width=0.7\textwidth]{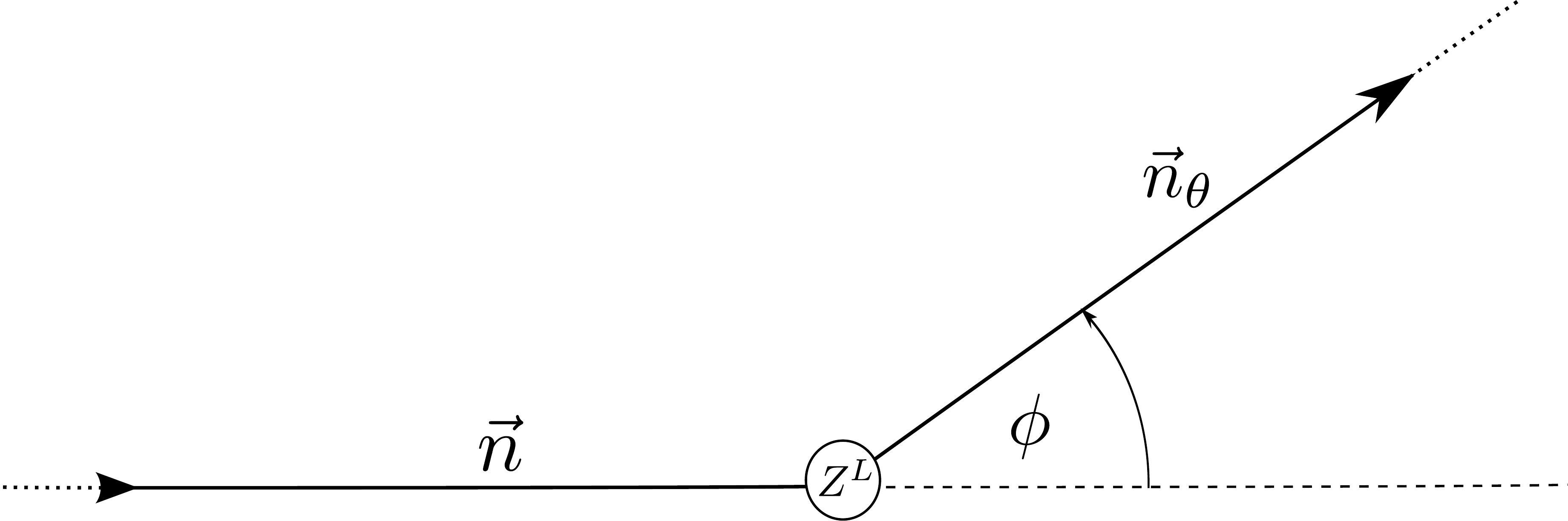}
	\caption{The cusped Wilson line with an operator insertion.}
	\label{fig:WL}
\end{figure}

The observable which we will be considering is the same as in \cite{Gromov:2012eu},\cite{Correa:2012hh} and \cite{Drukker:2012de}: it consists of two rays of a supersymmetric Wilson line forming a cusp with the angle $\phi$ and an operator $Z^L$ inserted at the cusp, where $Z$ is a scalar of ${\cal N}=4$ SYM (see figure \ref{fig:WL}). To completely define a supersymmetric Wilson line we should also specify the coupling to scalars, which is parameterized by a six-dimensional unit vector $\vec n(t)$ at each point of the line ($t$ being a parameter on the line). In our case $\vec n(t)$ is constant and equal $\vec n$ on one ray and $\vec n_\theta$ on another ray, so that $\vec n\cdot \vec n_\theta=\cos\theta$. Due to the R-symmetry the observable depends on $\vec n,\vec n_\theta$ only through $\theta$.
 Explicitly the observable is defined as
\beq
\label{WilsL}
	W_L={\rm P}\exp\!\int\limits_{-\infty}^0\! dt\(i  A\cdot\dot{x}_q+\vec\Phi\cdot\vec n\,|\dot x_q|\)\times Z^L\times {\rm P}\exp\!\int\limits_0^\infty\!dt\(i A\cdot\dot x_{\bar q}+\vec\Phi\cdot\vec n_\theta\,|\dot x_{\bar q}|\).
\eeq
Due to the cusp the expectation value of such an observable diverges as
\beq
\left\langle W_L\right\rangle \sim \(\frac{\Lambda_{IR}}{\Lambda_{UV}}\)^{\Gamma_L(\lambda)},
\eeq
where $\Lambda_{IR}$ and $\Lambda_{UV}$ are the infra-red and ultraviolet cut-offs respectively \cite{Polyakov:1980ca},\cite{Correa:2012at}. The quantity $\Gamma_{L}$, which we will call the cusp anomalous dimension, will be the main object of our studies.
When $\theta^2-\phi^2=0$ the observable $W_L$ becomes BPS and the cusp anomalous dimension vanishes \cite{Drukker:2006xg}. In \cite{Gromov:2012eu} the anomalous dimension in the near-BPS limit $\theta=0$, $\phi\rightarrow 0$ was calculated. In \cite{Gromov:2013qga} the calculation was generalized to the case of arbitrary, but close to each other angles $\theta^2-\phi^2\rightarrow 0$. In this general case the cusp anomalous dimension was found to be
\beq
\label{eq:mainresultIntro}
\Gamma_L(\lambda)=\frac{\phi-\theta}{4}\partial_\theta\log\frac{\det{\cal M}_{2L+1}}{\det{\cal M}_{2L-1}}+{\cal O}((\phi-\theta)^2),
\eeq
where ${\cal M}_N$ is an $(N+1)\times (N+1)$ sized matrix defined as
\begin{align}
\label{eq:M}
&\({\cal M}_N\)_{ij} =I^\theta_{i-j+1},\\
&I_n^\theta =i^{n+1}I_n\(\frac{\sqrt{\lambda}}{\sin\beta}\)\sin{n\beta}, \;\;\;\; \mathrm{with} \;\; \sin{\beta}=\frac{1}{\sqrt{1-\theta^2/\pi^2}},
\nonumber
\end{align}
 and $I_n(x)$ are modified Bessel functions of the first kind.
The AdS/CFT duality allows one to relate the configuration of ${\cal N}=4$ SYM fields described above to an open string in AdS which ends on a cusped line on the boundary of AdS. In particular, in the classical scaling limit when $L$ and $\lambda$ are both taken to infinity with $L/\sqrt{\lambda}$ fixed, we can match $\Gamma_{L}$ with the energy of the classical string. However, since the result contains determinants of $(2L+1)\times (2L+1)$ sized matrices it is not obvious how to take large  $L$ limit. In the subsequent sections we develop the apparatus for this, describe the classical string solution and finally compare the results for the energy.


\newpage
\section{Matrix model reformulation}
\label{sec:Matrix}

Taking the classical limit $L \rightarrow \infty$ keeping $L/\sqrt{\lambda}$ fixed becomes considerably easier once we realize that the cusp anomalous dimension \eqref{eq:mainresultIntro} can be expressed in terms of an expectation value of some operator in a matrix model. In this section we will show how to use this approach to find the large $N$ expansion of the determinant of ${\cal M}_N$ defined in \eqref{eq:M}.

One can check that the quantities $I^\theta_n$ defined in the previous section can be rewritten in the following integral representation
\begin{equation}
	I_n^\theta = \frac{1}{2\pi i} \oint \frac{dx}{x^{n+1}} \sinh(2\pi g \, (x + 1/x)) \, e^{2 g \theta (x - 1/x)},
\end{equation}
where the integration contour is the unit circle and $g=\frac{\sqrt\lambda}{4\pi}$. This makes it possible to write the determinant of ${\cal M}_{N}$ as
\beq
\det \M_{N}=
\oint\prod_{i=1}^{N+1}\frac{dx_i}{2\pi i} e^{2 g \, \theta \left(x_i-\frac{1}{x_i}\right)} \sinh \left(2 \pi g \left( x_i + \frac{1}{x_i} \right) \right) \times \det X,
\eeq
where
\beq
\det X = \left|
\bea{lllll}
x_1^{-2}&x_1^{-1}&\dots&x_1^{N-1}&x_1^{N-2}\\
x_2^{-3}&x_2^{-2}&\dots&x_2^{N}&x_2^{N-1}\\
\vdots&\vdots& \ddots & \vdots& \vdots\\
x_{N}^{-N-1}&x_{N}^{-N}&\dots&x_{N}^{-2}&x_{N}^{-1}\\
x_{N+1}^{-N-2}&x_{N+1}^{-N-1}&\dots&x_{N+1}^{-3}&x_{N+1}^{-2}
\eea
\right| = \frac{\prod_{i<j}^{N+1} (x_i - x_j)}{\prod_{i=1}^{N+1} x_i^{i+1}},
\eeq
and we recognize the numerator as the Vandermonde determinant $\Delta(x_i)$. We can further simplify the final result by anti-symmetrizing the denominator, which we can do because everything else in the integrand is anti-symmetric and the integration measure is symmetric w.r.t $x_i$, thus under the integral we can replace $\det X$ by
\beq
	\det X' = \frac{\Delta^2(x_i)}{(N+1)!} \, \prod_{i=1}^{N+1} \frac{1}{x_i^{N+2}}.
\eeq
Thus finally we get the following expression
\begin{equation}
  \det \M_{N} = \frac{1}{(2 \pi i)^{N+1}} \oint \prod_{i=1}^{N+1} \frac{d x_i}{x_i^{N+2}} \, \frac{\Delta^2(x_i)}{(N+1)!}  \, \sinh(2\pi g \, (x_i + 1/x_i)) \, e^{2 g \theta (x_i - 1/x_i)},
  \label{eq:Mintegral}
\end{equation}
which indeed has the structure of a partition function of some matrix model\footnote{Namely, it is equal to the partition function of a two-matrix model. We thank I.Kostov for discussions related to this question.}. It now becomes a matter of simple algebra to convince oneself that the cusp anomalous dimension \eqref{eq:mainresultIntro} can be written in terms of expectation values in this matrix model, namely
\beq
	\Gamma_L(g) = g \, \frac{\phi-\theta}{2} \left[ \,\, \left< \sum_{i=1}^{2L+1} \( x_i - \frac{1}{x_i}\)  \right>_{2L+1} - \,\,\, \left<  \sum_{i=1}^{2L-1} \(x_i - \frac{1}{x_i}\) \right>_{2L-1} \right],
\eeq
where $\left< \dots \right>_{N}$ denotes the normalized expectation value in the matrix model of size $N$ with the partition function defined in (\ref{eq:Mintegral}). Note that this formula is exact and we have not yet taken any limits.


\subsection{Saddle point equations}

In this section we will explore the classical $L\sim\sqrt \lambda\rightarrow \infty$ limit of the matrix model \eqref{eq:Mintegral}.  As usual in matrix models, when the size of matrices becomes large, the partition function is dominated by the solution of the saddle point equations. In the leading order it is just equal to the value of the integrand at the saddle point. Here we work in this approximation, leaving the corrections (beyond the first one calculated in section \ref{sec:The1loop}) for future work.

The partition function \eqref{eq:Mintegral}  can be recast in the form \footnote{we take $N=2L$.}
\beq
	\det \M_{2L} = \frac{1}{(2 \pi i)^{2L+1}} \frac{1}{(2L+1)!} \, \oint \prod_{i=1}^{2L+1} d x_i \, e^{-S(x_1, x_2, \dots, x_{2L+1})},
\eeq
where the action is given by
\begin{eqnarray}
	S &=& \sum_{i=1}^{2L+1} \[ 2 g \theta  \( x_i - \frac{1}{x_i} \)  - \(2L + 2\) \log x_i \] + 2 \sum_{i<j}^{2L+1} \log(x_i - x_j) + \\ \nonumber
	       &+& \sum_{i=1}^{2L+1} \log \sinh \( 2 \pi g \( x_i + \frac{1}{x_i} \) \).
\end{eqnarray}
The saddle point equations $\partial S / \partial x_j = 0$ now read\footnote{Technically the $x_j^{-1}$ term has a coefficient of $L+1$, but since we are taking $L \rightarrow \infty$ we chose to neglect it for simplicity.}
\begin{equation}
	g \theta \left( 1 + \frac{1}{x_j^2} \right) - \frac{L}{x_j} + \sum_{i \neq j}^{2L+1} \frac{1}{x_j - x_i} + \pi g \left( 1- \frac{1}{x_j^2} \right) \coth \left( 2 \pi g \left( x_j + \frac{1}{x_j} \right) \right) = 0.
\end{equation}
We can further simplify them by noting that a large coupling constant $g$ appears inside the cotangent and since the roots $x_i$ are expected to be of order 1, with the exponential precision
it is possible to replace
\begin{equation}
	\coth \left( 2 \pi g \left( x_j + \frac{1}{x_j} \right) \right) \approx \mathrm{sgn}(\mathrm{Re}(x_j)).
\end{equation}
Finally we bring the equations to a more canonical and convenient form and get the following result,
\begin{equation}
	-\theta \, \frac{x_j^2 + 1}{x^2_j - 1} + \frac{L}{g} \frac{x_j}{x_j^2 - 1} - \frac{1}{g} \frac{x_j^2}{x_j^2 - 1} \sum_{i \neq j}^{2L+1} \frac{1}{x_j - x_i} = \pi \, \mathrm{sgn}(\mathrm{Re}(x_j)).
	\label{eq:saddlepoint}
\end{equation}
An alternative way of finding these values $x_i$ is to consider the following quantity $P_L(x)$, which played an important role in \cite{Gromov:2012eu},
\beq
P_L(x)=\frac{1}{\det {\cal M}_{2L}}\left|\begin{matrix}
I_1^{\theta}& I_0^{\theta}& \cdots & I_{2-2L}^{\theta}  &I_{1-2L}^{\theta}\\
I_2^{\theta}& I_1^{\theta}& \cdots & I_{3-2L}^{\theta} &I_{2-2L}^{\theta}\\
\vdots      &  \vdots     &\ddots & \vdots            &\vdots           \\
I_{2L}^{\theta}& I_{2L-1}^{\theta}& \cdots & I_{1}^{\theta}  &I_{0}^{\theta}\\
x^{-L}& x^{1-L}& \cdots & x^{L-1} &x^{L}\\
\end{matrix}\right|.
\label{eq:PLrepr}
\eeq
The numerator is the same as $\det \M_{2L}$ except in the last line $x_{2L+1}$ is replaced by $x$ which is not integrated over. In the classical limit all integrals are saturated by their saddle point values, i.e. one can remove the integrals by simply replacing $x_i \rightarrow x_i^{cl}$. If we replace $x$ with any saddle point value $x_i^{cl}$ the determinant will contain two identical rows and will automatically become zero, thus the zeros of $P_L(x)$ are the saddle point values. On the complex plane they are distributed on two arcs as shown in figure \ref{fig:roots}. As expected, for the case $\theta=0$ we recover two symmetric arcs on the unit circle \cite{Gromov:2012eu}.

Now, following \cite{Gromov:2007aq},\cite{Beisert:2005bm},\cite{Gromov:2012eu}, we introduce the quasimomentum $p(x)$ as
\begin{equation}
	p(x) = -\theta \, \frac{x^2 + 1}{x^2 - 1} + \frac{L}{g} \frac{x}{x^2 - 1} - \frac{2L}{g} \frac{x^2}{x^2 - 1} \, G_L(x),
	\label{eq:quantum_quasimomentum}
\end{equation}
where the resolvent $G_L(x)$ is
\begin{equation}
	G_L(x) = \frac{1}{2L} \sum_{k=1}^{2L+1} \frac{1}{x-x_k}.
	\label{eq:resolvent}
\end{equation}
The motivation for introducing $p(x)$ is that  the saddle point equations (\ref{eq:saddlepoint}) expressed through $p(x)$ take a very simple form
\begin{equation}
	\frac{1}{2} \left( p(x_i + i \epsilon) + p(x_i - i \epsilon) \right) = \pi \, \mathrm{sgn}(\mathrm{Re}(x_i)).
	\label{eq:pepsilon}
\end{equation}
In the classical limit the poles in the quasimomentum condense and form two cuts. The shifts $\pm i\epsilon$ in the equation above refer to taking the argument of the quasimomentum to one or the other side of the cut.


\subsection{Closed form expression for the quasimomentum}

The quasimomentum \eqref{eq:quantum_quasimomentum} introduced in the previous section is a convenient object to consider when taking the classical limit $L\sim \sqrt \lambda\rightarrow\infty$, because in this limit it is related to the algebraic curve of the corresponding classical solution. In this section we will construct this curve explicitly.

\begin{figure}[t]
	\centering
		\includegraphics[width=0.95\textwidth]{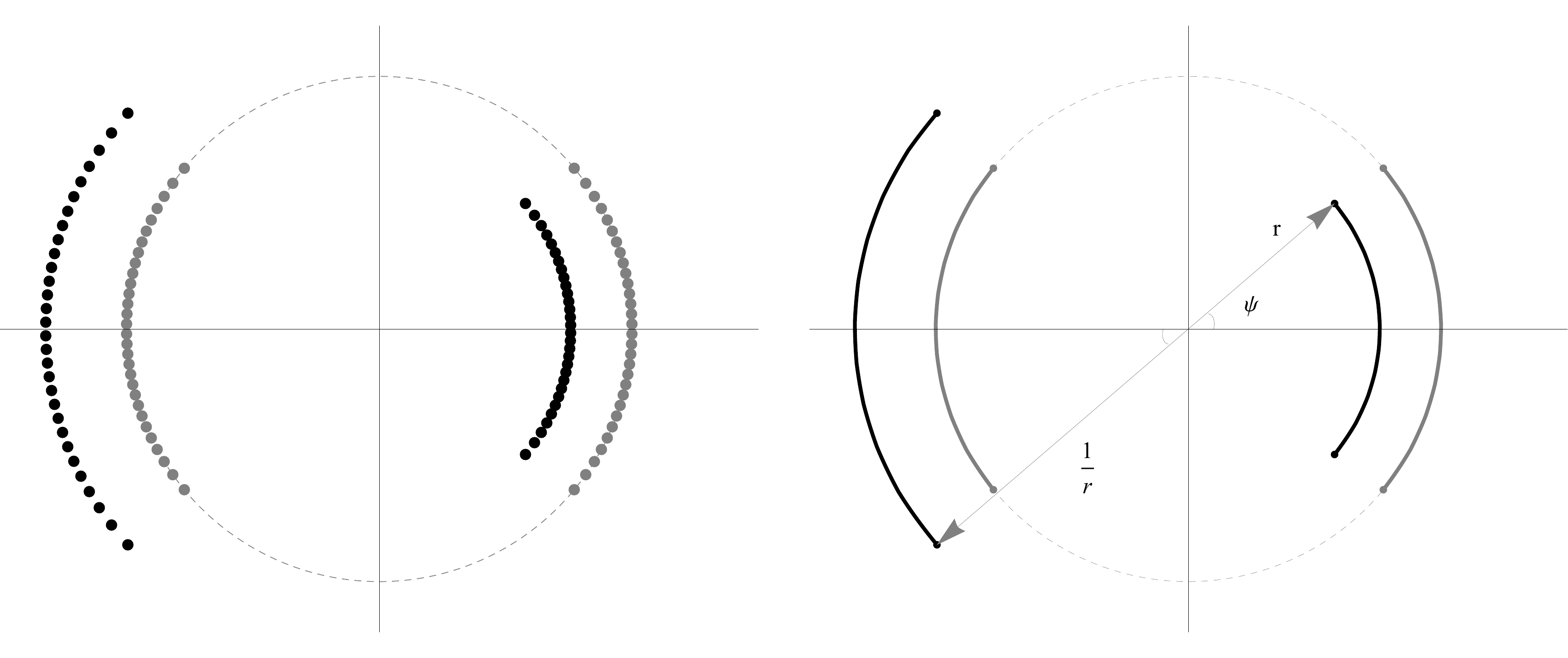}
	\caption{Distribution of roots on the complex plane for $\theta=0$ (gray) and $\theta=1$ (black) on the left and the condensation of the roots to corresponding smooth cuts on the right with the algebraic curve parameters $r$ and $\psi$ identified. The dashed circle is the unit circle.}
	\label{fig:roots}
\end{figure}

In the classical limit the poles of $p(x)$, which we denote as $x_i$, are governed by the saddle-point equation and condense on two cuts in the complex plane, as shown in figure \ref{fig:roots}.
The saddle-point equation \eqref{eq:saddlepoint} has a symmetry $x\rightarrow -1/x$, so does the set of poles $x_i$. For the quasimomentum (\ref{eq:quantum_quasimomentum}) this symmetry manifests as the identity $p(x) = -p(-1/x)$. Thus we conclude that the two cuts are related by an $x\rightarrow -1/x$ transformation. This and the invariance of the saddle-point equation under complex conjugation implies that the four branch points can be parameterized as $\{r\,e^{i\psi},r \, e^{-i\psi},-1/r \, e^{i\psi},-1/r \, e^{-i\psi}\}$. Note that in the case $\theta=0$ the symmetry is enhanced to $p(x)=-p(-x)$ and $p(1/x) = p(x)$, which is not true for arbitrary $\theta$.

The crucial point to notice is that while $p(x)$ satisfies the equation \eqref{eq:pepsilon} which has different constants on the right hand side for the two different cuts, the corresponding equation for $p'(x)$ has a zero on the right hand side for both cuts, thus we expect $p'(x)$ to have a simpler form than $p(x)$. Our strategy is to write down an ansatz for the derivative $p'(x)$ using the symmetries and analytical properties of $p(x)$ and then integrate it. The form of the expression we get is analogous to the curve constructed in \cite{Kazakov:2004qf}, which also helps us to construct the ansatz.

First, $p(x)$ has four branch points and according to \eqref{eq:pepsilon} its derivative changes sign on each cut, hence all the cuts are of square-root type. One can write $p'(x)\propto 1/y(x)$, where
 \beq
 	y(x) = \sqrt{x-r e^{i \psi}}\sqrt{x-r e^{-i \psi}}\sqrt{x + \frac{1}{r} e^{i \psi}}\sqrt{x + \frac{1}{r} e^{-i \psi}}.
 \eeq
  Second, since the algebraic curve is obtained from \eqref{eq:quantum_quasimomentum} in the classical limit, $p(x)$ should have simple poles at $x=\pm 1$. Finally, from \eqref{eq:quantum_quasimomentum}  we can get the behaviour at infinity:
\beq
	p'(x) \approx \frac{L}{g} \frac{1}{x^2} + \mathcal{O}\(\frac{1}{x^3}\).
\eeq
By using the knowledge about these singularities and asymptotics we can fix $p(x)$ completely. Based on what we know up to now we write down our ansatz for the derivative
\beq
	p'(x) = \frac{A_1 x^4 + A_2 x^3 + A_3 x^2 + A_4 x + A_5}{(x^2 - 1)^2 \, \sqrt{x-r e^{i \psi}}\sqrt{x-r e^{-i \psi}}\sqrt{x + \frac{1}{r} e^{i \psi}}\sqrt{x + \frac{1}{r} e^{-i \psi}}}.
\eeq
The polynomial in the numerator is of order four in order to maintain the correct asymptotics, and below we fix its coefficients using the properties of the quasimomentum.\footnote{Comparing with the asymptotic one can immediately see that $A_1 = L/g$, however our objective is to express $p(x)$ solely in terms of $r$ and $\psi$, which parameterize the algebraic curve.}

 The $x\rightarrow-1/x$ symmetry for the derivative implies that $A_1 = A_5$ and $A_2 = -A_4$. Next, simple poles at $x=\pm 1$ in $p(x)$ require zero residues of $p'(x)$ at $x=\pm1$, which fixes $A_2$ to be
\beq
	A_2 =- \frac{(2A_1 + A_3) \, r \, (r^2 - 1) \cos \psi}{r^4 - 2 \, r^2 \, \cos 2\psi + 1}.
\eeq
We fix the two remaining unknowns $A_1$ and $A_3$ after integrating the $p'(x)$. We don't write the intermediate results of the integration as the expressions are enormous without any apparent structure.
Looking back at \eqref{eq:pepsilon} we see that at the branchpoints
\beq
	p(x_{bp}) =\pm \pi.
\eeq
We use this condition to fix $A_1$ and we get

\beq
	A_1 = \frac{A_3}{2}\frac{K_1-E_1}{E_1+K_1-2\,a^2 \, K_1\cos^2(\psi)},
\eeq
where
\beq
E_1=\mathbb{E}\(a^2\sin^2(\psi)\),\;K_1=\mathbb{K}\(a^2\sin^2(\psi)\),\;a = \frac{2r}{r^2+1}.
\label{eq:E1K1}
\eeq
Finally we can use the $x \rightarrow -1/x$ symmetry on the quasimomentum itself, as before we only used it on the derivative. Imposing the symmetry yields

\beq
	A_3 = \frac{8}{a}\(E_1+K_1-2\,a^2\,\cos^2(\psi)K_1\).
\eeq
As expected, after plugging  these coefficients into $p(x)$ (and using the identities from the Appendix \ref{sec:Elliptic}) the whole expression simplifies enormously and we are left with our main result
\begin{align}
	\label{eq:pmainresult}
	p(x) &= \pi - 4\,i\,  E_1\, \mathbb{F}_1(x) + 4\,i\,  K_1\, \mathbb{F}_2(x) - a \left( \frac{x+ r e^{-i\psi}}{x+\frac{1}{r} e^{i\psi}} \right) \left(\frac{2/r\,e^{i \psi}}{x^2 - 1} \right) y(x)\,K_1,
\end{align}
where
\begin{equation}
	\mathbb{F}_1(x) = \mathbb{F}\left( \left. \sin^{-1} \sqrt{ \left( \frac{x -r e^{-i\psi}}{x+ \frac{1}{r} e^{i\psi}} \right) \left( \frac{ e^{ i \psi}}{ia\,r\sin\psi} \right)} \; \right| a^2 \sin^2(\psi) \right),
\end{equation}
\begin{equation}
	\mathbb{F}_2(x) = \mathbb{E}\left( \left. \sin^{-1} \sqrt{ \left( \frac{x -r e^{-i\psi}}{x+ \frac{1}{r} e^{i\psi}} \right) \left( \frac{ e^{ i \psi}}{ia\,r\sin\psi} \right) } \; \right| a^2 \sin^2(\psi) \right).
\end{equation}
We verified this result numerically by comparing it to the extrapolation of the discrete quasimomentum \eqref{eq:quantum_quasimomentum} at large $L$ and got an agreement up to thirty digits. We also compared this expression at $\theta = 0$ with the quasimomentum obtained in \cite{Gromov:2012eu} and the expressions agree perfectly.

The resulting quasimomentum is parameterized in terms of the branchpoints, i.e. the parameters are the radius $r$ and angle $\psi$. They are determined in terms of $L/g$ and $\theta$, which are parameters of the matrix model. We already mentioned that $L/g$ is simply the constant $A_1$, which we found to be

\beq
	\frac{L}{g} = 4\,\frac{K_1-E_1}{a} ,
	\label{eq:Lgfixed}
\eeq
and looking back at \eqref{eq:quantum_quasimomentum} we see that $\theta = p(0) = -p(\infty)$, hence
\beqa
	\theta &=& -\pi + \frac{2a}{r}\,e^{i\psi}K_1 \nonumber \\
	       &-& \left. 4 \, i \, K_1\,\mathbb{E}\left(\sin ^{-1}\sqrt{\frac{  e^{i\psi}}{ia\,r\sin\psi}} \,\right|\, a^2 \sin ^2(\psi )\right)\nonumber \\
           &+& \left. 4 \, i \, E_1\, \mathbb{F}\left(\sin ^{-1}\sqrt{\frac{  e^{i\psi}}{ia\,r\sin\psi}}\,\right|\, a^2 \sin ^2(\psi )\right).
           \label{eq:thetafixed}
\eeqa

In the next section the two equations above will be matched with two analogous equations following from the classical string equations of motion.


\section{Classical string solution}
\label{sec:Classical}

As we have mentioned before, in the classical $L\sim\sqrt{\lambda}\rightarrow\infty$ limit $\Gamma_L(\lambda)$ can be matched with the energy of an open string. In this section we will describe the corresponding string solution and find the classical energy.

The class of string solutions we are interested in was introduced in \cite{Correa:2012hh} and generalized in \cite{Gromov:2012eu}. It is a string in $AdS_3\times S^3$ governed by the parameters $\theta,\phi$, $AdS_3$ charge $E$ and $S^3$ charge $L$; the four parameters are restricted by the Virasoro constraint. The ansatz for the embedding coordinates of $AdS^3$ and $S^3$ is
\begin{align}
y_1+iy_2=e^{i\kappa\tau}\sqrt{1+r^2(\sigma)},\;\; y_3+iy_4=r(\sigma) e^{i\phi(\sigma)},\\
x_1+ix_2=e^{i\gamma\tau}\sqrt{1+\rho^2(\sigma)},\;\; x_3+ix_4=r(\sigma) e^{i f(\sigma)}.
\end{align}
The range of the worldsheet coordinate is $-s/2<\sigma<s/2$, where $s$ is to be found dynamically. The angles $\theta$ and $ \phi$ parameterizing the cusp enter the string solution through the boundary conditions $\phi(\pm s/2)=\pm (\pi-\phi)/2$ and $f(\pm s/2)=\pm\theta/2$. The equations of motion and Virasoro constraints lead to the following system of equations (see Appendix E of \cite{Gromov:2012eu} for more details, also \cite{Drukker:2011za}):
\begin{align}
f(\gamma,l_\theta)&=f(\kappa,l_{\phi}),
\label{eq:ff}
\\
h(\gamma,l_{\theta})=\theta, &\;\; h(\kappa,l_{\phi})=\phi,
\label{eq:hh}
\\
g(\gamma,l_\theta)=L, &\;\; g(\kappa,l_\phi)=E,\label{eq:gg}
\end{align}
where
\begin{align}
f(\gamma,l)&=\frac{2\sqrt{2}}{\sqrt{\gamma^2+k^2+1}}\,\mathbb{K}\left(\frac{-k^2+\gamma^2+1}{k^2+\gamma^2+1}\right),
\label{eq:f}
\\
h(\gamma,l)&=\frac{2l}{k(1+k^2-\gamma^2)}\left[(1+\gamma^2+k^2)\,\Pi\left(\frac{k^2-2l^2-\gamma^2+1}{2k^2}\,\vline\,\frac{k^2-\gamma^2-1}{2k^2}\right)-\right.
\nonumber
\\ & \left.-2\gamma^2\,\mathbb{K}\left(\frac{k^2-\gamma^2-1}{2k^2}\right)\right],
\label{eq:h}
\\
g(\gamma,l)&=-2\sqrt{2} \, \frac{\sqrt{\gamma^2+k^2+1}}{\gamma}\left[\mathbb{E}\left(\frac{-k^2+\gamma^2+1}{k^2+\gamma^2+1}
\right)-\mathbb{K}\left(\frac{-k^2+\gamma^2+1}{k^2+\gamma^2+1}
\right)\right],
\label{eq:g}
\\
k^4&=\gamma^4-2 \gamma^2+ 4\, \gamma^2 l^2+1.
\nonumber
\end{align}

One can see that the variables $\theta,l_{\theta},\gamma$ and $L$ are responsible for the $S^3$ part of the solution, while  $\phi,l_{\phi},\kappa$ and $E$ are their analogues for $AdS_3$. The two parts of the solution are connected only by the Virasoro condition which leads to \eqref{eq:ff}.
We are interested in the limit when $\theta\approx\phi$. In this limit the two groups of variables responsible for $S^3$ and $AdS_3$ parts of the solution become close to each other, namely $l_{\theta}\approx l_{\phi}$ and $E\approx L$. The cusp anomalous dimension should be compared with the difference $E-L$, because $L$ is the classical part of the dimension of the observable $W_L$. To find $E-L$ we linearize the system \eqref{eq:f},\eqref{eq:h},\eqref{eq:g} around $\phi\approx\theta$, which yields
\begin{align}
E-L=(\phi-\theta)\left|\frac{\partial{(g,f)}}{\partial{(l,\kappa)}}\right|/\left|\frac{\partial{(h,f)}}{\partial{(l,\kappa)}}\right|.
\label{eq:ELbig}
\end{align}
Plugging in here the explicit form of $g,f$ and $h$ one gets as a result an extremely complicated expression with a lot of elliptic functions. However, there exists a parametrization in which the result looks surprisingly simple: this parametrization comes from comparison of the string conserved charges with the corresponding quantities of the algebraic curve. One can notice that the equations for $\theta$ and $L/g$ in the end of the last section have the same structure as the equations \eqref{eq:hh} and \eqref{eq:gg}. Indeed, it is possible to match them precisely if one chooses the correct identification of parameters of the string solution $l_\theta,\gamma$ with the parameters of the algebraic curve $r,\psi$. We used the elliptic identities presented in the appendix \ref{sec:Elliptic} to bring the equations to identical form after the following identifications
\beq
\gamma=-\frac{2r}{\sqrt{r^4-2r^2\cos2\psi+1}},\; l_{\theta}=\frac{(r^2-1)\cos\psi}{\sqrt{r^4-2r^2\cos2\psi+1}}.
\eeq
As another confirmation of correctness of this identification, after plugging it into \eqref{eq:ELbig} the complicated expression reduces to the following simple formula for the classical energy
\begin{align}
E-L=g(\phi-\theta)(r-1/r)\cos\psi.
\label{eq:E1}
\end{align}
Notice that this can be rewritten as a sum over the branch points of the algebraic curve
\begin{align}
&E-L=\frac{g}{2}(\phi-\theta)\sum\limits_i a_i,
\end{align}
where $a_i=\{r \, e^{i\psi}, r \, e^{-i\psi},-1/r \, e^{i\psi},-1/r \, e^{-i\psi}\}$.


\section{The energy from the quasimomentum}
\label{sec:Theenergy}

In this section we will find the classical limit of the cusp anomalous dimension from the algebraic curve. At large $L$ the formula \eqref{eq:mainresultIntro} can be rewritten as
\beq
\Gamma_L(g)=\frac{\phi-\theta}{4}\partial_\theta\partial_L\det{\cal M}_{2L}.
\eeq
Use the integral representation \eqref{eq:Mintegral} for $\det{\cal M}_L$ we can notice that
\beq
\partial_\theta \log \det {\cal M}_L=\left\langle 2g\sum\limits_{i=1}^{2L}(x_i-1/x_i)\right\rangle,
\eeq
where by the angular brackets we denoted an expectation value in the matrix model with the partition function \eqref{eq:Mintegral}.
In the quasiclassical approximation the expectation value is determined by the saddle-point, i.e. the previous expression is equal to  $2g\sum\limits_{i=1}^{2L}(x_i-1/x_i)$, where the roots $x_i$ are the solutions of the saddle-point equation \eqref{eq:saddlepoint}.
Since the set of the roots has a $x\rightarrow-1/x$ symmetry, the two terms in the sum give the same contribution. Thus
\beq
\partial_\theta \log \det {\cal M}_L=-4g \sum\limits_{i=1}^{2L}\frac{1}{x_i}=8 \, g \, L \, G(0),
\eeq
where we used the resolvent \eqref{eq:resolvent}.

Using the relation \eqref{eq:quantum_quasimomentum} between the resolvent and the quasimomentum  we find $G(0)=\frac{g}{L}\left(p''(0)/4-\theta\right)$, so the final expression for the cusp anomalous dimension in terms of the quasimomentum is
\begin{align}
\Gamma_L(g)=-\frac{g^2}{2}\partial_Lp_L''(0).
\label{eq:E2}
\end{align}
The formula for $p(x)$ presented in the previous section is given in terms of the parameters of the branch points $r$ and $\psi$. They are implicitly defined through $L/g$ and $\theta$ by the equations \eqref{eq:Lgfixed} and \eqref{eq:thetafixed}. In order to get $\Gamma_L$ we express $\partial_L$ though $\partial_r$ and $\partial_\psi$ and then apply \eqref{eq:E2} to \eqref{eq:pmainresult}. Finally we obtain a very simple result in terms of $r$ and $\psi$
\beq
\Gamma_{L}(g)=g(\phi-\theta)\(r-1/r\)\cos\psi
\label{eq:GammaL}
\eeq
which exactly coincides with the calculation from the string solution!


\subsection{Comparison with the small angle limit}

Here we will check our formula \eqref{eq:GammaL} in the limit $\phi=0$ and $\theta\rightarrow 0$ considered in section E.2 of \cite{Gromov:2012eu}. As the angles go to zero, the branch points approach the unit circle: $r\rightarrow 1$, thus the formula \eqref{eq:GammaL} gives
\beq
	\Gamma_L(g)=2\,g\,\theta (r-1)\cos\psi.
\eeq
In this limit $r-1\propto\theta$, and the coefficient of proportionality can be found by expanding\footnote{The equation \eqref{eq:thetafixed} is written in the approximation $\phi\approx\theta$ and now on the top of it we want to take a limit $\theta\rightarrow 0$. Since before we have neglected the terms ${\cal O}(\theta-\phi)^2$, the result, which is now of the order ${\cal O}(\theta)^2$ will not generally be reproduced. However, we found that here and in several other formulas correct small angle limit is reproduced if before taking $\theta,\phi$ to zero we replace $\theta$ and $\phi$ by the middle angle $\phi_0=\frac{\phi+\theta}{2}$, which is in our case equal to $\theta/2$.} the equation \eqref{eq:thetafixed} for $\theta$ around $r=1$:
\beq
	2 (1-r)\frac{\mathbb{E}\left(\sin^2\psi\right)}{\cos\psi}=\theta/2.
\eeq
Plugging it into the formula above we get
\beq
\Gamma_L(g)=g\,\theta^2 \,\frac{\cos^2\psi}{2\mathbb{E}\left(\sin^2\psi\right)}
\eeq
which perfectly agrees with (190) of \cite{Gromov:2012eu}.

\subsection{The 1-loop correction to the classical energy}
\label{sec:The1loop}
Now that the classical limit of the cusp anomalous dimension is calculated, we can consider corrections to it. In the limit $L\sim \sqrt\lambda\rightarrow\infty$ which we are studying here
the perturbative expansion around the classical value can be written as
\beq
\Gamma_L(g)=\sum\limits_{n=0}^\infty g^{1-n}b_n(L/g)+\text{non-perturbative terms}.
\label{eq:Gammaexp}
\eeq
The classical energy is $g \, b_0(L/g)$ and other corrections are suppressed by powers of $g$. A symmetry of the formula for $\Gamma_L(g)$ found in \cite{Beccaria:2013lca} allows one to express the even terms in the expansion \eqref{eq:Gammaexp} through the odd ones and the other way round. In particular, $b_1$ can be obtained from $b_0$ by differentiating with respect to $L/g$.
Since the classical energy is
\beq
\Gamma^{cl}_L(g)=g\(\phi-\theta\)\(r-1/r\)\cos\psi
\eeq
by differentiating it with respect to $L/g$ we find that the perturbative part of energy in the first two orders in the classical expansion is
\begin{equation}
\Gamma_L(g)=g\(\phi-\theta\)\(r-1/r\)\cos\psi \left(1+\frac{1}{g}f(r,\psi)\right),
\label{eq:correctedenergy}
\end{equation}
where
\begin{align}
f(r,\psi)=\frac{r+1/r}{4}\frac{\left|r^2e^{2i\psi}+1\right|^2K_1-r^2 \left|r+\frac{1}{r}+e^{i\psi}-e^{-i\psi}\right|^2E_1 }{\left|\(r+\frac{1}{r}\)\(r^2e^{2i\psi}-1\)E_1-\(r-\frac{1}{r}\)\(r^2e^{2i\psi}+1\)K_1\right|^2},
\end{align}
and $E_1,K_1$ are defined in \eqref{eq:E1K1}. We have checked this formula and the classical energy \eqref{eq:E1} against a numerical extrapolation of the exact expression \eqref{eq:mainresultIntro} and found an agreement up to more than thirty digits.


\section{Conclusions}
\label{sec:Conclusions}

In this note we considered the cusped Wilson line operator studied in \cite{Gromov:2013qga}. We presented a matrix model formulation of the result obtained in \cite{Gromov:2013qga} for the anomalous dimension of the Wilson line, which is convenient when exploring the classical limit $L\sim \sqrt{\lambda}\rightarrow\infty$. We found the corresponding classical algebraic curve \eqref{eq:pmainresult} and derived a simple formula for the energy of the dual classical string solution \eqref{eq:E1}. We also calculated the classical energy from the algebraic curve and verified numerically that those two expressions match with $\Gamma_{L}$ given by \eqref{eq:mainresultIntro} in the classical limit.
In \cite{Beccaria:2013lca} an important observation about the expansion of $\Gamma_{L}$ around the classical solution was made, which is that the expansion is fully determined by half of the coefficients, i.e. the odd coefficients can be calculated from the even ones and the other way round. Based on this and our knowledge of the classical energy we calculated here the 1-loop correction to the classical energy \eqref{eq:correctedenergy}.

The natural way to proceed exploring the properties of the cusp anomalous dimension at strong coupling is by studying
the algebraic curve. The algebraic curve was found here by taking the classical limit of the quantized expression \eqref{eq:quantum_quasimomentum}, but the properties of the result are identical to those of the finite-gap solutions \cite{Beisert:2005bm}. This is quite surprising, because here we are dealing with open strings, for which the finite-gap procedure is not yet developed and a priori one could expect new features of the curve, for example, contributions from the boundary. It would be interesting to generalize the finite-gap method to the open string case and obtain the quasimomentum \eqref{eq:pmainresult} directly from the classical solution.

 Based on the algebraic curve presented here, it would be interesting to generalize it to the case of arbitrary $\theta$ and $\phi$. This may give a possibility to find the expansion of $\Gamma_{L}$ around the classical solution away from the near-BPS limit.


\section*{Acknowledgements}

We are grateful to I.Kostov, N.Gromov and F.Levkovich-Maslyuk for helpful discussions and comments on the manuscript. We thank N.Gromov for formulating the problem, supervising the work and helping at all stages of the project. We acknowledge the support of the GATIS network and thank IPhT Saclay for hospitality. The research leading to these results has received funding from the People Programme (Marie Curie Actions) of the European Union’s Seventh Framework Programme FP7/2007-2013/ under REA Grant Agreement No 317089.


\newpage
\appendix

\section{Elliptic Identities}
\label{sec:Elliptic}

This appendix contains the identities involving elliptic functions which we used to simplify expressions throughout the paper. For a real $z$
\begin{align}
\frac{\mathbb{E}(z)}{\sqrt{1-z}}&=\begin{cases}
i\mathbb{E}\(\frac{z}{z-1}\),\;z<1,\\
\mathbb{E}\(\frac{z}{z-1}\)+2i\[\mathbb{K}\(\frac{1}{1-z}\)-\mathbb{E}\(\frac{1}{1-z}\)\],\;z>1.
\end{cases}\\
\sqrt{1-z}\mathbb{K}(z)&=\begin{cases}
\mathbb{K}\(\frac{z}{z-1}\),\;z<1,\\
\mathbb{K}\(\frac{z}{z-1}\)+2i\mathbb{K}\(\frac{1}{1-z}\),\;z>1.
\end{cases}
\end{align}
The following two-parametric identity holds for $r>0,\;0<\psi<\pi/2$:
\begin{align}
\pi&=\frac{4r^2}{r^2+1} \, e^{i\psi} \,\mathbb{K}(\sin^2(q))\\ \nonumber
&+\frac{4r}{r^2-1}\tan q \, \cos\psi \left[\mathbb{K}\(-\tan^2(q)\)-\frac{r^2+1}{4r^2}\Pi\(\frac{(r^2-1)^2}{4r^2}\tan^2(q) \, \vline\, -\tan^2(q)\)\right]\\ \nonumber
&+4i\[\mathbb{E}\(\sin^2(q)\)\mathbb{F}\(\sin^{-1}\(\sqrt{\frac{r^2+1}{2}}\sqrt{1-i\cot\psi}\)\,\vline\,\sin^2(q)\)\right.\\ \nonumber
&\left.-\mathbb{K}\(\sin^2(q)\)\mathbb{E}\(\sin^{-1}\(\sqrt{\frac{r^2+1}{2}}\sqrt{1-i\cot\psi}\)\,\vline\,\sin^2(q)\)\],
\end{align}
where $\sin^2(q)=\frac{4 r^2\sin^2\psi}{(r^2+1)^2}$.


\end{document}